# Numerical Aspects of Gradient Reconstruction Schemes Applied to Complex Geometries

Frederico Bolsoni Oliveira[1*] and João Luiz F. Azevedo[2]

[1*]Instituto Tecnológico de Aeronáutica, São José dos Campos, 12228-900, São Paulo, Brazil.
[2]Instituto de Aeronáutica e Espaço, São José dos Campos, 12228-904, São Paulo, Brazil.

*Corresponding author(s). E-mail(s): fredericobolsoni@gmail.com;
Contributing authors: joaoluiz.azevedo@gmail.com;

**Abstract**

The present work primarily focuses on the study of three gradient reconstruction techniques applied to the calculation of viscous terms in a cell-centered, finite volume formulation for general unstructured grids. The work also addresses different ways of formulating the limiter functions necessary to maintain stability in the presence of flow discontinuities. The flows of interest are simulated using the compressible Reynolds-averaged Navier-Stokes equations, and the negative Spalart-Allmaras model is used for turbulence closure. Definition of interface inviscid terms uses the Roe approximate Riemann solver, whereas the interface viscous terms are calculated with a standard centered scheme together with appropriate definitions of the interface gradients. Steady state solutions are obtained using an implicit time-integration method, together with a novel convergence acceleration technique. This new approach defines a set of three simple rules for controlling the global CFL number based on the residue evolution. The work considers three test cases, namely, the subsonic bump-in-channel flow, the subsonic NASA high-lift Common Research Model multielement airfoil and the transonic ONERA M6 wing. Present results are compared to experimental and numerical data available in the literature. Severe numerical instabilities are observed when the simplest gradient reconstruction technique is used, while more sophisticated formulations are able to provide excellent agreement with the existing literature. Current results are demonstrated to be highly insensitive to modifications made to the entropy fix terms of the numerical flux. Integrated aerodynamic forces are shown to be mildly dependent on the limiter formulation used, even in the absence of shock waves. The proposed convergence acceleration procedure manages to quickly drive the residue terms to machine zero, provided no major instabilities are present.

## 1 Introduction

For many decades, numerical simulations of physical models have been used to aid the design of engineering solutions to real-life problems. The numerical data obtained from these simulations can provide the user with significant insights into the fundamental physics of the problem being analyzed, which would otherwise be impractical or prohibitively expensive to obtain experimentally [1]. Unfortunately, the use of numerical schemes to discretize the equations of interest typically introduces unwanted error terms to the original model [2], thus disturbing the solution and possibly reducing the physical correlation of the obtained data. The influence of these additional error terms on the numerical solution can be mitigated by either improving the quality of the discretization or by making use of numerical operators that better approximate the original equation [3, 4]. Therefore, any improvements made to the formulation of those numerical operators are certainly welcomed by the scientific and engineering communities.

In the context of computational fluid dynamics (CFD), in particular the subset that deals with high-Reynolds number compressible flows, a very common discretization procedure applied to the gas dynamics equations is the finite volume method (FV), coupled with a piecewise continuous solution reconstruction. For cell-centered FV methods, a key aspect that must be carefully addressed is the calculation of discrete properties, and their gradients, at cell interfaces. Such values must be appropriately reconstructed from cell-averaged values preferably using compact stencils, thus allowing for efficient implementation in massively parallel computing environments. The present study is especially interested in addressing the calculation of these interface property gradients, which are necessary for the calculation of interface viscous fluxes [5–7], when used in conjunction with general unstructured grids.

Among all strategies considered by the authors, a common step in the calculation of the interface property gradients is the evaluation of the cell-averaged property gradient. In that regard, two approaches dominate the unstructured, FV literature, namely the Green-Gauss [5, 8–11] and the least-square [11–15] methods. The former extends the original FV discrete property definition to also include the property gradients. That is, the cell property gradients are treated as being the cell-averaged property gradients. This way, one can leverage the Green-Gauss theorem to effectively compute the desired gradients by performing a series of property reconstructions at each cell interface. This approach can be very precise when applied to homogeneous meshes, particularly structured ones, but may produce numerical disturbances and solution degradation in regions of the domain in which different cell geometry types are present [16, 17]. The latter, appropriately named the least-squares method, tries to avoid the main weakness of the Green-Gauss scheme by locally approximating the solution by a linear function, extending that function to the direct neighbors of the current cell and minimizing the error that exists between this extended function and the average property values of each cell. The approach is less sensitive to mesh topology changes, but its effectiveness is highly dependent on the proper definition of weighting functions employed in the process of minimizing the error [13, 16]. Furthermore, depending on the implementation, the least-squares method has the potential of becoming significantly more computationally expensive than the Green-Gauss method. Alternative approaches do exist, but they either employ a hybrid formulation [17, 18], in which both the Green-Gauss and the least-squares methods are combined in an attempt to address their shortcomings, or apply modifications to the original methods [9, 10, 19], typically performed to reduce the mesh influence on the solution.

Once the cell property gradients are computed, the face property gradients can be determined by adopting one of the many schemes available in the literature, as listed in Refs. [5, 7, 20, 21]. Nevertheless, no matter which scheme is chosen, the face property gradient is determined by some type of average of the direct adjacent cell property gradients, followed by a sequence of correction functions that try to improve some aspect of the discrete operator, as shown later in this paper. Unfortunately, most of the analysis regarding the behavior of these face gradient reconstruction techniques that are available in the literature involve artificially created conditions, such as the use of highly disturbed meshes and simplified case configurations, that are purposely designed to highlight the differences that exist between each scheme [5, 7, 21]. Thus, the current effort attempts to remedy this issue by analyzing three different face gradient reconstruction techniques when applied to the solution of simplified test cases that hold some resemblance to typical configurations faced by a CFD engineer on a day-to-day basis. The objectives of the research effort include providing numerical data that could allow a better understanding of the influence of the different gradient reconstruction approaches, in particular for the case of compressible turbulent flows.

The present effort considers the flows of interest are adequately described by the compressible Reynolds-averaged Navier-Stokes (RANS) equations [4]. Turbulence closure is achieved with the negative Spalart-Allmaras (SA-neg) model [22–24]. Equations are discretized by a cell centered, finite volume method for general unstructured meshes [25–27]. Cell interface properties are calculated by a limited, piecewise linear reconstruction of primitive variables, yielding a 2nd-order accurate, total-variation diminishing (TVD) scheme [8, 28]. The overall research effort tested different limiters [29, 30], but all the results included here have used the Wang limiter [20], which is a modification of the Venkatakrishnan limiter [31]. Inviscid cell interface fluxes are calculated using the Roe approximate Riemann solver with entropy fix [26, 27, 32]. Viscous fluxes at cell interfaces are computed by a standard centered discretization scheme. Three different interface gradient calculation schemes are implemented, which use average cell gradients computed with a weighted Green-Gauss formulation [26, 27]. All results shown here are steady-state calculations. Hence, an implicit Euler method is used for the time march, together with a GMRES [33] linear system solver. A novel process for dynamically controlling the global CFL number at given time step is proposed, and the results demonstrate that it yields very stable and nearly optimal convergence rates. The details that distinguish the proposed approach from other methods currently available in the literature are strictly technical and are presented in Sec. 2.3. The main contributions of the present work are associated with the studies of different cell-interface gradient reconstruction methods, the analysis of appropriate limiters for property reconstruction, and the formulation of an efficient convergence acceleration procedure. Moreover, the present calculations also aim to demonstrate that adequately formulated steady-state drivers can yield tightly converged solutions, even for fairly complex physical problems.

The work addresses three different test cases. The first one is the three-dimensional, subsonic bump-in-channel flow [24], solved on regular hexahedral meshes. The second test case is the NASA high-lift common research model (CRM-HL) multielement airfoil [24, 34, 35]. The final test case is the three-dimensional ONERA M6 wing [24, 36]. The paper describes the details of the formulation previously outlined, together with a careful discussion of the results obtained for these test cases. The formulation has been implemented into an in-house developed code [25–27], which is used for all calculations presented here. It is hoped that the present results can contribute to a better understanding of the effects of the various numerical aspects here addressed in the calculation of compressible turbulent flows.

## 2 Theoretical and Numerical Formulation

In the present study, the system of conservation laws of interest, *i.e.*, the RANS equations [4] together with the SA-neg turbulence model [22–24], is numerically solved using a general unstructured grid, cell-centered, finite volume method. The current section presents the general finite volume framework and the mathematical formulation of each gradient reconstruction scheme examined in this study, along with the details of the time marching scheme.

### 2.1 Governing Equations and Overall Numerical Formulation

The governing equations can be written as a system of conservation laws as

$$\frac{\partial \vec{Q}}{\partial t} + \vec{\nabla} \cdot \vec{\mathcal{F}}(\vec{Q}, \overrightarrow{\nabla Q}) = 0 \ , \tag{1}$$

where $\vec{Q}$ is the vector of conserved quantities, $t$ is the time variable, and $\vec{\mathcal{F}}$ is a geometric vector of algebraic flux vectors. The flux can be separated into the contributions of the inviscid and viscous terms as

$$\vec{\mathcal{F}} \equiv \vec{\mathcal{F}}_e - \vec{\mathcal{F}}_v \ . \tag{2}$$

This would allow rewriting Eq. (1) as

$$\frac{\partial \vec{Q}}{\partial t} + \vec{\nabla} \cdot \left[ \vec{\mathcal{F}}_e(\vec{Q}) - \vec{\mathcal{F}}_v(\vec{Q}, \overrightarrow{\nabla Q}) \right] = 0 \ , \tag{3}$$

In a 3-D Cartesian system, the flux vectors could be explicitly written as

$$\vec{\mathcal{F}}_e(\vec{Q}) \equiv \vec{E}_e(\vec{Q}) \, \hat{i} + \vec{F}_e(\vec{Q}) \, \hat{j} + \vec{G}_e(\vec{Q}) \, \hat{k} \ , \tag{4}$$

and

$$\vec{\mathcal{F}}_v(\vec{Q}, \overrightarrow{\nabla Q}) \equiv \vec{E}_v(\vec{Q}, \overrightarrow{\nabla Q})\,\hat{i} + \vec{F}_v(\vec{Q}, \overrightarrow{\nabla Q})\,\hat{j} + \vec{G}_v(\vec{Q}, \overrightarrow{\nabla Q})\,\hat{k}\ . \qquad (5)$$

Clearly, the $e$ and $v$ subscripts are used to refer to the inviscid and viscous components of the flux vectors, respectively. Moreover, $\hat{i}$, $\hat{j}$ and $\hat{k}$ are the triad of unitary vectors aligned with the $x$, $y$ and $z$ coordinate directions, respectively. The functional relation that exists between each flux vector and $\vec{Q}$ and $\overrightarrow{\nabla Q}$ has been explicitly stated in Eq. (3). In this fashion, the portions of the equation affected by each property reconstruction can be easily identified. Further details in the definition of the flux vectors can be found in the literature [3, 25–27].

As it is usual with the finite volume approach, Eq. (1) can be integrated over control volumes which are obtained by dividing the computational domain into multiple discrete cells of polyhedral shape. Hence, performing the integration of a generic cell of volume $\mathbb{V}$, and using the divergence theorem to handle the flux terms, the governing equations can be rewritten in integral form as

$$\frac{\partial}{\partial t} \int_{\mathbb{V}} \vec{Q}\, d\mathbb{V} + \oint_S \vec{\mathcal{F}} \cdot \overrightarrow{dS} = 0\ , \qquad (6)$$

where $S$ is the outer surface of the cell with volume $\mathbb{V}$, and the vector elemental area $\overrightarrow{dS}$ is always pointing outwards in the normal direction. Equation (6) must be true for all cells that compose the domain. Hence, by defining $n_f$ as the number of faces of the $i$-th cell and by employing a single point Gaussian quadrature, the formulation can be written in a space-discretized form as

$$\mathbb{V}_i \frac{\partial \vec{Q}_i}{\partial t} + \sum_{k=1}^{n_f} \left( \vec{\mathcal{F}}_k \cdot \vec{S}_k \right) = 0\ , \qquad (7)$$

where the discrete vector of conserved properties at the $i$-th volume, $\vec{Q}_i$, is defined as

$$\vec{Q}_i \equiv \frac{1}{\mathbb{V}_i} \int_{\mathbb{V}_i} \vec{Q}\, d\mathbb{V}\ . \qquad (8)$$

Moreover, $\vec{\mathcal{F}}_k$ represents the flux vector reconstructed at the $k$-th face of the $i$-th cell. The approach used here for writing the spatial discretization of the governing equations is valid up to 2nd-order accurate schemes, which is all that we are dealing with in the present paper.

In order to solve Eq. (7), it is necessary to define a numerical flux to approximate the inviscid and viscous components of the face reconstructed flux vector, $\vec{\mathcal{F}}_k$. The reconstructed inviscid flux uses the Roe approximate Riemann solver [26, 27, 32] in order to formulate the numerical flux. Hence, it can be written as

$$\vec{\mathcal{F}}_{e_k} \cdot \vec{S}_k = \vec{\mathcal{F}}\left(\vec{Q}_k\right) - \frac{1}{2}\left|\tilde{A}_k\right|\left|\vec{S}_k\right|\left(\vec{Q}_{kj} - \vec{Q}_{ki}\right)\ , \qquad (9)$$

where $\left|\tilde{A}_k\right|$ is the Roe matrix formed with the magnitude of the characteristic speeds, and associated with the $k$-th face normal direction. Moreover,

$$\vec{Q}_k = \frac{1}{2}\left(\vec{Q}_{ki} + \vec{Q}_{kj}\right)\ , \qquad (10)$$

where $i$ and $j$ are the indices of two adjacent cells. Terms $\vec{Q}_{ki}$ and $\vec{Q}_{kj}$ are vectors of conserved variables written as a function of $\vec{V}_{ki}$ and $\vec{V}_{kj}$, respectively. Properties $\vec{V}_{ki}$ and $\vec{V}_{kj}$ are the $i$-th and $j$-th cell reconstructed vectors of primitive variables, respectively, evaluated at the centroid of the $k$-th face. For a 2nd-order TVD scheme, a limited, piecewise linear reconstruction [8, 28] is used, which can be written as

$$\vec{V}_{ki} = \vec{V}_i + \psi_i \nabla \vec{V}_i \cdot \vec{r}_{ki}\ , \qquad (11)$$

$$\vec{V}_{kj} = \vec{V}_j + \psi_j \nabla \vec{V}_j \cdot \vec{r}_{kj}\ . \qquad (12)$$

In Eq. (11), $\vec{r}_{ki}$ is the distance vector that points from the centroid of the $i$-th cell to the centroid of the $k$-th face. Equivalent definition is also used for $\vec{r}_{kj}$ in Eq.(12). The limiter, $\psi$, is computed separately for each variable of interest.

In the present work, Wang's modification to the Venkatakrishnan limiter [20, 31] is used. The Venkatakrishnan limiter [31] is defined as

$$\psi^{VK}\left(\frac{\Delta_+}{\Delta_-}\right) = \frac{\left(\Delta_+^2 + \epsilon^2\right) + 2\Delta_-\Delta_+}{\Delta_+^2 + 2\Delta_-^2 + \Delta_+\Delta_- + \epsilon^2} \ , \quad (13)$$

where $\Delta_+$ and $\Delta_-$ are property variations measured from the current cell and its immediately adjacent cells, that is, those cells that share a common face with the $i$-th cell. These variations are defined as

$$\Delta_- \equiv \overline{V}_{ki} - V_i \ , \quad (14)$$

$$\Delta_+ = \begin{cases} V_{\max} - V_i, & \text{if } \overline{V}_{ki} \geq V_i \ , \\ V_{\min} - V_i, & \text{if } \overline{V}_{ki} < V_i \ , \end{cases} \quad (15)$$

where

$$\overline{V}_{ki} \equiv V_i + \overrightarrow{\nabla V_i} \cdot \vec{r}_{ki} \ . \quad (16)$$

Moreover, $V_{\max}$ and $V_{\min}$, in Eq. (15), are the maximum and minimum discrete primitive property values among the current cell and all of its immediate neighbors. The $\epsilon^2$ parameter is a modification to the standard definition of a limiter, introduced by Venkatakrishnan [37], to avoid numerical complications in nearly constant regions. In the original definition of the limiter, it is made proportional to a mesh length scale, $\Delta x$ [37]. However, such a definition can introduce computational problems, since meshes can have very large variations in cell size. Hence, values of $\epsilon^2$ can vary significantly along the mesh simply due to the differences in the length scales associated to the largest and smallest cells in the computational domain. Hence, the work of Wang [20] attempts to correct this issue by making $\epsilon$ proportional to flow property variations. This is, therefore, the formulation adopted here, which writes

$$\epsilon = \epsilon_W \left(V_{\text{gl.max}} - V_{\text{gl.min}}\right) \ . \quad (17)$$

In Eq. (17), $\epsilon_W$ is a constant whose recommended values lie in the range $\epsilon_W \in [0.01, 0.20]$, and $V_{\text{gl.max}}$ and $V_{\text{gl.min}}$ are, respectively, the global maximum and minimum values of the corresponding property being limited over the entire computational domain.

It is an usual practice, in the implementation of the Roe numerical flux, to create additional artificial dissipation at the point in which the eigenvalues of the Roe matrix change sign. This procedure is typically called an entropy fix formulation in the literature [38]. Hence, the calculation of the absolute value of the Roe matrix eigenvalues, $|\lambda|$, when evaluating the $\left|\tilde{A}_k\right|$ matrix in Eq. (9), adds an additional term that avoids zero artificial dissipation when the eigenvalue changes sign [39]. This modification can be written as

$$\left.|\lambda|\right|_{fix} = \begin{cases} \frac{1}{2}\left(\frac{\lambda^2}{\epsilon_H} + \epsilon_H\right), & \text{if } |\lambda| < \epsilon_H, \\ |\lambda|, & \text{otherwise.} \end{cases} \quad (18)$$

When the formulation is appropriately set in a dimensionless form, the recommended values [38, 39] for $\epsilon_H$ lie within the range $0 \leq \epsilon_H \leq 0.25$.

The spatial discretization of the viscous fluxes uses a centered approach. Hence, common interface properties used to compute the viscous terms are calculated using simple averages, as indicated in Eq. (10). The calculation of the viscous fluxes, however, also required common cell interface gradients. The construction of such cell interface property gradients is discussed in the next subsection.

## 2.2 Gradient Reconstruction Schemes

Three gradient reconstruction schemes are used in the present effort, and these are denoted $L00$, $L0E$ and $LJ0$. The labelling convention adopted here follows the work of Jalali and co-authors [7] and it is formed by a sequence of three characters. The 1st character indicates the type of weighted-averaging which is performed in order to obtain the face gradient from the corresponding cell-averaged gradients. In this case, $L$ stands for a linear reconstruction. The 2nd character indicates if some type of jump construct is used [6]. Finally, the 3rd character is associated to the use of any further correction procedure. In the case of the $L0E$ scheme, for instance, $E$ stands for an edge-based correction scheme. These three schemes were chosen due to their mathematical simplicity, potential to provide good results, as shown in Ref. [7], reasonable computational cost and compatibility with the current programming framework built by the authors. The formulation of each of these gradient reconstruction schemes, as implemented in the present work, is described next.

### 2.2.1 L00 Scheme

The $L00$ gradient reconstruction scheme is one of the simplest formulations possible. In this scheme, the cell interface property gradient is calculated as the weighted average of the two directly adjacent average cell-gradient values. The averaging process is weighted by the distances between the adjacent cell centroids and the current face centroid. Hence, the interface gradient of a generic $A$ scalar quantity, evaluated at the $k$-th face shared by the $i$ and $j$ cells, can be written as

$$(\overrightarrow{\nabla A})_k = \frac{|\vec{r}_{kj}|\,(\overrightarrow{\nabla A})_i + |\vec{r}_{ki}|\,(\overrightarrow{\nabla A})_j}{|\vec{r}_{ki}| + |\vec{r}_{kj}|} \equiv (\overline{\nabla A})_k \;. \tag{19}$$

Here, it is convenient to define $(\overline{\nabla A})_k$ as the $L00$ interface gradient at the $k$-face for the sake of writing the expressions for the other gradient reconstruction procedures which will be discussed in the forthcoming paragraphs. Moreover, the average cell gradients, $(\overrightarrow{\nabla A})_i$ and $(\overrightarrow{\nabla A})_j$, are computed using a weighted Green-Gauss (GG) approach, where the weights follow the same rationale to what has been introduced in Eq. (19). For instance, the average cell gradient for the $i$-th cell is defined as

$$\left(\overrightarrow{\nabla A}\right)_i \equiv \frac{1}{\mathbb{V}_i} \sum_{k=1}^{n_f} \left( \frac{|\vec{r}_{kj}|\,A_i + |\vec{r}_{ki}|\,A_j}{|\vec{r}_{ki}| + |\vec{r}_{kj}|} \right) \vec{S}_k \;. \tag{20}$$

An equivalent definition is used for $\left(\overrightarrow{\nabla A}\right)_j$. The authors obviously acknowledge that other definitions for the average cell gradients exist in the literature [12, 16]. However, the present work has used the fairly simple definition indicated in Eq. (20).

It is well known in the literature [40, 41] that, although quite simple and inexpensive to compute, the $L00$ scheme yields a computational stencil that effectively does not use information from the $i$ and $j$ cells. Hence, high-frequency errors can develop in the solution with the use of such interface gradient calculation procedure [7]. A possible approach to attempt to remedy this condition is to augmented the $L00$ scheme formulation with the introduction of extra terms that ensure dependency on cell-averaged data of the two cells that share the interface. The $L0E$ and $LJ0$ schemes are doing precisely that, by looking at the issue from two different perspectives.

### 2.2.2 L0E Scheme

The $L0E$ scheme, also known as the edge-normal scheme, consists in modifying the interface gradient calculated by the $L00$ scheme by exchanging the gradient component in the direction that connects the $i$ and $j$ cell centroids with a straightforward finite difference construct [7, 40]. Hence, considering the geometric definitions illustrated in Fig. 1, the formulation of $L0E$ interface gradient reconstruction scheme can be written as

$$(\overrightarrow{\nabla A})_k = (\overline{\nabla A})_k + \left[\frac{A_j - A_i}{|\vec{r}_{ij}|} - (\overline{\nabla A})_k \cdot \frac{\vec{r}_{ij}}{|\vec{r}_{ij}|}\right] \frac{\vec{r}_{ij}}{|\vec{r}_{ij}|} \, , \tag{21}$$

where $(\overline{\nabla A})_k$ is the interface gradient obtained from the $L00$ scheme, as previously defined. As indicated in Fig. 1, $\vec{r}_{ij}$ is the vector connecting the centroid of the $i$-th cell to that of the $j$-th cell. Hence, this approach essentially reintroduces the contributions of the $i$ and $j$ cells into the stencil that computes $(\overrightarrow{\nabla A})_k$ in the $L0E$ scheme.

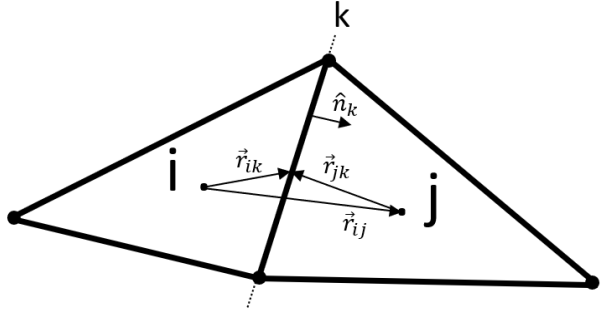

**Fig. 1** Mesh schematic diagram picturing cells $i$ and $j$. Focus is given to the $k$-th face of the $i$-th cell. Face normal unitary vector, $\hat{n}_k$, as well as relevant distance vectors, $\vec{r}$, are also shown. The dots are used to represent the cell centroid locations.

### 2.2.3 LJ0 Scheme

The $LJ0$ scheme incorporates a jump term construct, in which information from the discontinuous nature of the solution at the face centroid is introduced in the formulation of the interface gradient reconstruction [6, 7]. In this context, the expression for the cell interface gradient of a generic $A$ scalar quantity can be written as

$$(\overrightarrow{\nabla A})_k = (\overline{\nabla A})_k + \frac{\alpha}{|\vec{r}_{ij} \cdot \hat{n}_k|} \left(A_{k_j} - A_{k_i}\right) \hat{n}_k \, . \tag{22}$$

Here, $A_{k_i}$ and $A_{k_j}$ are the piecewise linearly reconstructed $A$ properties of cells $i$ and $j$, respectively, evaluated at the interface centroid, using a formulation as presented in Eq. (16). Moreover, as indicated in Fig. 1, $\hat{n}_k$ is the face normal unit vector pointing outwards with respect to the current cell, i.e., the $i$-th cell. For the present implementation, the jump coefficient, $\alpha$, of the $LJ0$ scheme is taken as $\alpha = 4/3$. This choice is based on the recommendation in the literature [6], and it is associated with guaranteeing a maximum possible order of accuracy for the overall numerical scheme.

The literature [6] also shows that different interface gradient reconstruction techniques can be cast into the form of Eq. (22). For instance, the $L0E$ scheme, previously presented, can also be written in the form of Eq. (22) with the following definition for the $\alpha$ coefficient:

$$\alpha = (\hat{n}_k \cdot \hat{e}_{ij}) |\hat{n}_k \cdot \hat{e}_{ij}| \, , \tag{23}$$

where

$$\hat{e}_{ij} \equiv \frac{\vec{r}_{ij}}{|\vec{r}_{ij}|} \tag{24}$$

is the unit vector in the direction connecting the centroids of the $i$-th and $j$-th cells.

### 2.3 Time-Integration and Convergence Acceleration Scheme

In the present effort, steady state solutions are obtained by integrating the equations using a first-order, implicit Euler scheme. An important issue for the implementation of implicit schemes in a general unstructured grid, parallel environment is the construction of the Jacobian matrices, in particular for the representation of the viscous terms. Typically, in such computational environments, it is not efficient to represent the complete Jacobian matrix, due to the number of cells that would be involved in the computational stencil, directly affecting the compactness of such stencil. Hence, simplifications must be made. On the other hand, if too many simplifications are made, this can have a significant impact in the convergence rates and, in some cases, even in the stability of the convergence process. In the present implementation, the authors use approximate analytically-derived Jacobian matrices that represent the most relevant terms, particularly for the viscous operator and the turbulence model. Clearly, all the inviscid terms are completely represented in the Jacobian matrix. The present approach yields a compact stencil in the construction of the linear systems that must be solved at each time step.

Solutions to the linear systems are obtained by using the Restarted Generalized Minimum Residue (GMRES(m)) iterative method, in which the Arnoldi process is restarted after $m$ consecutive iterations [33]. Here, $m$ is taken to be equal to 200 for all calculations presented. Due to the ill-conditioned nature of these linear systems, a preconditioner is used. Since the parallel implementation of the present code works mainly with distributed memory, the Additive Schwarz [42] preconditioner is applied to the entire linear system, in conjunction with an Incomplete Lower-Upper factorization with fill level 3, ILU(3), applied locally to each partition of the domain, as recommended in the literature [43, 44].

The present work proposes a simple and robust, but very efficient, process to dynamically increase the simulation Courant-Friedrichs-Lewy (CFL) throughout the calculation, that has demonstrated to yield convergence to machine zero for most of the test cases analyzed. Hence, for all simulations performed in the present work, the CFL number is dynamically increased from 0.1 to 10,000, based on the current residue behavior, as measured by an appropriate convergence sensor which will be described in the forthcoming paragraphs. The current algorithm is an original approach, which is, however, inspired by the ideas discussed by different authors in the literature [45–47]. The global CFL number is updated in each iteration by following a sequence of three different control conditions. In their most general form, these conditions are applied to the $L_\infty$-norm of the residue associated with each equation being solved. In the present effort, however, they are applied only to the $L_\infty$-norm of the residues associated with the continuity equation and the turbulence closure model, as the other residues display a similar behavior as compared to that presented by these two equations. For simplicity, the $L_\infty$-norm of the residue will be referred to simply as "residue" ($Res.$) throughout this subsection.

In all cases, the algorithm monitors not only the values of the residue in the current iteration ($n$), but also the values of the residue of the previous two iterations. Depending on the observed behavior, the algorithm can assume three different states, as follows:

- **The residue displays a monotone convergence behavior.** That is, if $Res.^{(n)} \leq Res.^{(n-1)} \leq Res.^{(n-2)}$, then the global CFL number is increased by 25%.
- **The computed conserved property value increment for the current time step, $\Delta \vec{Q}^{(n)}$, leads to a non-physical solution state.** In the current implementation, a non-physical solution state is achieved by having negative density and/or negative total energy anywhere in the domain. If this condition occurs, then the current non-linear iteration is discarded, *i.e.*, the calculated $\Delta \vec{Q}^{(n)}$ is not applied anywhere in the domain, and the CFL number is decreased by 98%.
- **The residue was decreasing, but the current iteration reversed that behavior.** That is, if $Res.^{(n-1)} \leq Res.^{(n-2)}$ and $Res.^{(n)} > Res.^{(n-1)}$, then $Res.^{(n-1)}$ is treated as a local minimum and it is stored as a lower reference level. An upper reference level is taken to be 10 times the lower reference level. While the residue stays in between these lower and upper reference levels, the CFL number is kept constant or it can be increased, or decreased, if the conditions stated in the previous two items are met. If the residue becomes higher than the previously defined upper reference level, then the global CFL number is decreased by 40% and both the upper and lower reference levels are removed from memory. Furthermore, if the residue crosses the lower reference level, then both the upper and lower reference levels are also removed from memory, without making any additional modification to the current CFL number, except for what is already prescribed in the previous two rules. When no reference level is stored, then the algorithm continues to monitor the state of the residue and the emergence of a new local minimum. This mechanism essentially allows the upper reference level to decrease during the convergence of the solution, always keeping it one order of magnitude higher than the previously registered local minimum.

These three conditions are applied simultaneously to the residues which are being monitored, which, in the present case, are the residues of the continuity equation and of turbulence closure model, as already indicated. The new CFL number is always taken to be the lowest value between the two calculated CFL values obtained by the previous process, by monitoring the evolution of the residues of the two equations. Although very simple and cheap to compute, this convergence control algorithm is seen to be capable of always keeping the current iteration CFL number to a reasonable value, without requiring any user input, for all cases studied here. Thus, nonlinear instabilities, that might arise when the solution process is started from freestream conditions with a sufficiently high CFL value, are avoided [44, 45].

## 3 Three-Dimensional Bump-in-Channel Flow

### 3.1 Description of the Test Case

The subsonic bump-in-channel flow is defined in NASA Langley's Turbulence Modeling Resource (TMR) website [24]. It is an interesting three-dimensional (3-D) test case, yet geometrically simple for an initial test of the present developments. The problem consists of a sinusoidal 3-D wall bump, located at the center of the bottom boundary of an otherwise inviscid channel. The bump area is modeled as a non-slip adiabatic wall, while the upstream and downstream regions of the bottom boundary are considered as inviscid symmetry planes. Likewise, the domain top, front and back boundaries are also symmetry planes. The left, or entry, boundary is a subsonic inlet with known stagnation properties. At the outflow boundary, the static pressure is set to its ambient value, here denoted $p_\infty$, with other properties extrapolated from the domain.

The "freestream" Reynolds number is set as $Re_\infty = 3 \times 10^6$, based on the reference length $L_{ref} = 1$ m and the reference temperature $T_\infty = 300$ K. At the inlet, it is assumed that the velocity direction is perpendicular to the boundary surface, and its magnitude is obtained from the interior of the domain using zero-order extrapolation. As indicated, stagnation properties are assumed known at the domain entrance, and they are obtained by defining that $P_t/p_\infty = 1.02828$ and $T_t/T_\infty = 1.008$, where $P_t$ and $T_t$ are the entrance stagnation pressure and temperature, respectively. Under these conditions, the entrance, or freestream, Mach number is about 0.2, which, therefore, characterizes a nearly incompressible test case that is being solved using a compressible formulation.

A family of three different hexahedral meshes with similar topology, but with different levels of refinement, is considered. Figure 2 shows an overview of mesh 3, with special focus on the region that surrounds the bump.

These hexahedral cell meshes are highly orthogonal, which typically result in a fairly benign test case. However, since the application of each scheme to this type of mesh still results in different numerical formulations, especially for the $L00$ scheme as discussed in Sec. 2.2.1, the authors believe it is an interesting case to consider in order to assess possible numerical anomalies that might arise in the discrete solution. The meshes are available in NASA Langley's TMR [24], and employs a cell distribution with a bias towards the near wall regions of the flow, as well as regions near the leading and trailing edges of the bump. In the present calculations, mesh 1 has 28,160 cells, mesh 2 has 225,280 cells and mesh 3 has 1,802,240 cells. Solutions obtained using the FUN3D and CFL3D codes, available in TMR [24], are used for comparison purposes. Besides the three meshes here described, results from FUN3D and CFL3D also employ an extra mesh, named mesh 4, composed of 14,417,920 hexahedral cells, which is not used in our calculations.

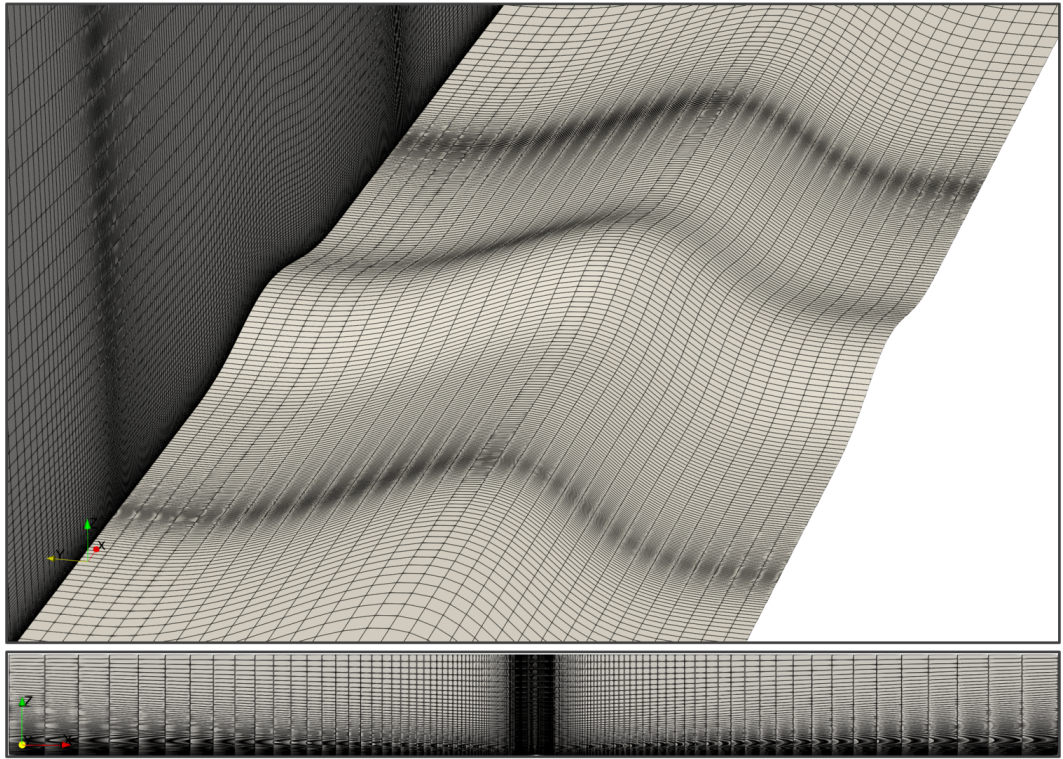

**Fig. 2** Overview of mesh 3 used in the bump-in-channel case. Focus is given to the region that surrounds the bump.

## 3.2 Discussion of Results – Effects of Gradient Reconstruction

The initial comparison of the present results is based on pressure coefficient distributions along the bump surface in the streamwise direction, $x$, at two different spanwise locations, namely $y = -0.50$ m and $y = -0.25$ m. The solution at $y = -0.50$ m is at the centerline of the computational domain. The results obtained with the present code, BRU3D, using the medium mesh, mesh 2, are shown in Fig. 3. Our results, with the three different interface gradient reconstruction schemes, are compared to those obtained with the NASA node-centered FUN3D code, calculated using mesh 4 [24]. There are essentially no differences among the various curves for the spanwise station at $y = -0.25$ m, whereas some differences can be observed at $y = -0.50$ m between the present results and those calculated with FUN3D. All three interface gradient reconstruction schemes implemented yield essentially the same results at the $y = -0.50$ m station too.

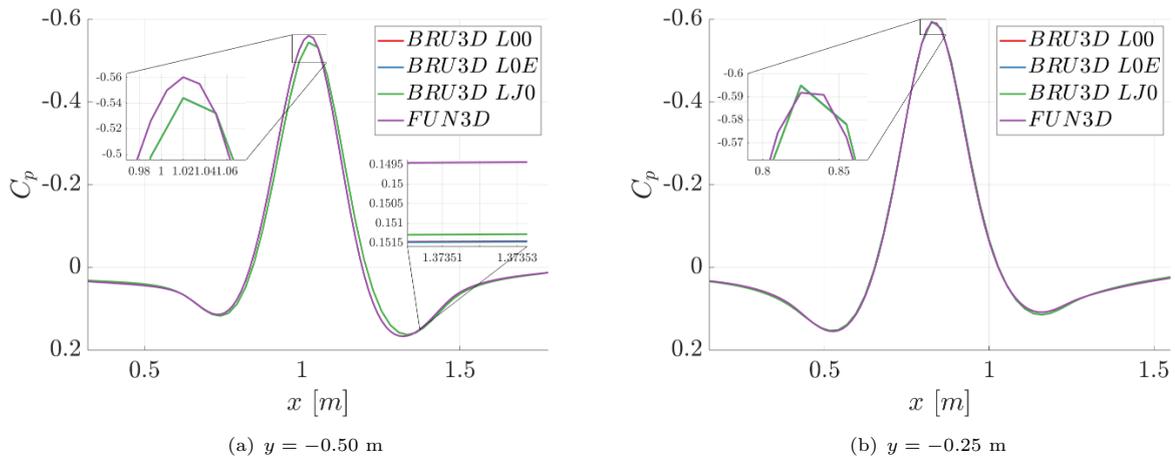

(a) $y = -0.50$ m  (b) $y = -0.25$ m

**Fig. 3** Pressure coefficient distributions along the bump surface obtained using the BRU3D code with mesh 2 and the FUN3D code with mesh 4.

Figure 4, which shows a comparison of the present results, computed using mesh 3, with the same FUN3D results [24], indicates a much better agreement even at the $y = -0.50$ m station. This comparison shows that the previously observed discrepancies were primarily due to differences in mesh refinement. Therefore, given a sufficiently fine mesh, the present calculations are essentially reproducing the results obtained with FUN3D in mesh 4. As already indicated, for this test case, the different gradient reconstruction techniques produced no discernible differences in the resulting pressure distributions over the bump. Although only two spanwise stations are shown here, similar results were obtained for pressure coefficient distributions along the entire computational domain in the streamwise direction.

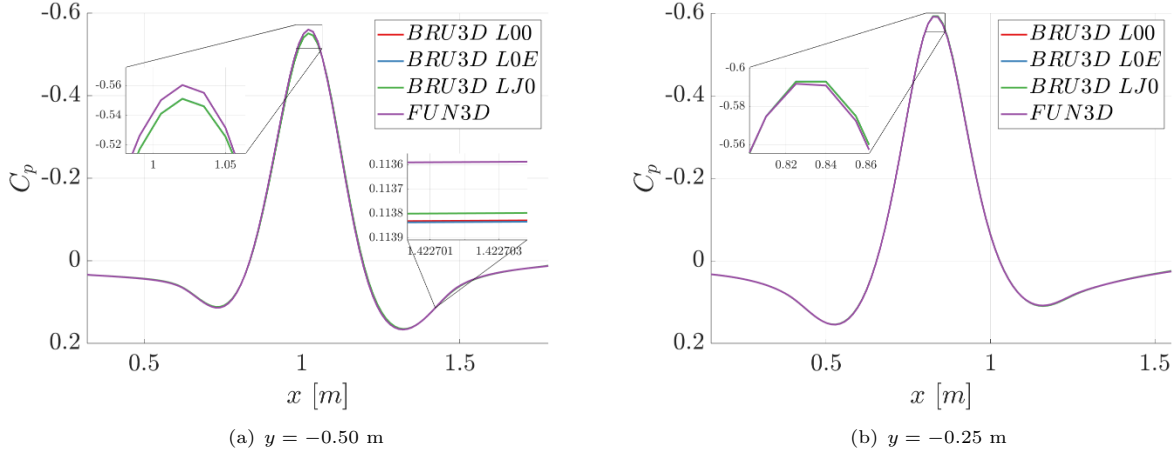

(a) $y = -0.50$ m

(b) $y = -0.25$ m

**Fig. 4** Pressure coefficient distributions along the bump surface obtained using the BRU3D code with mesh 3 and the FUN3D code with mesh 4.

Figure 5 presents results for the drag coefficient, $C_D$, as a function of the total number of cells, $N$, in the domain. The drag coefficient is computed over the surface of the bump, as indicated in the TMR website [24], and considers a reference area of 1.5 m$^2$. The present calculations are again compared to FUN3D data [24]. The results essentially indicate that there is a very small effect of the scheme for computing the interface gradients in the present calculations. Moreover, results also indicate that BRU3D and FUN3D calculations seem to be converging to the same solution, as the mesh refinement is increased. This behavior can be observed by the reduction in the gap between all curves in Fig. 5 with increased mesh refinement.

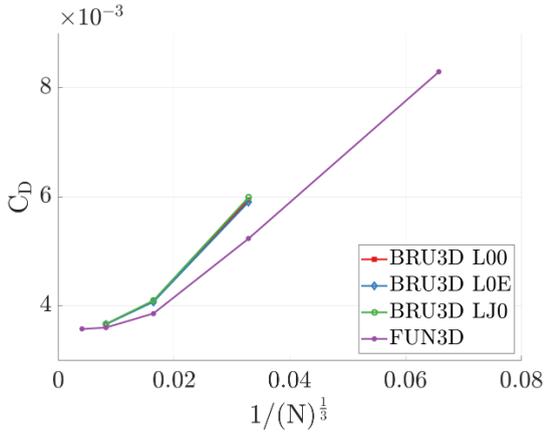
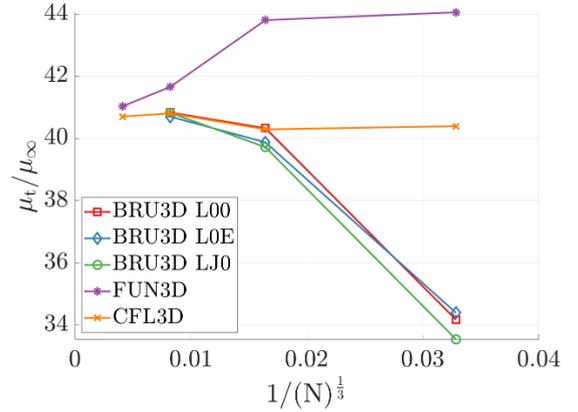

**Fig. 5** Drag coefficient, $C_D$, of the bump region computed using different levels of mesh refinement.

**Fig. 6** Dimensionless eddy viscosity, evaluated at $x = 0.3$ m, $y = -0.2$ m, $z = 0.0035$ m, using different levels of mesh refinement.

Values of dimensionless eddy viscosity, $\mu_t/\mu_\infty$, calculated by the SA-neg turbulence model at a particular point in the flowfield, are shown in Fig. 6 again as a function of mesh refinement. In this case, a randomly chosen point with coordinates $(x, y, z) = (0.3, -0.2, 0.0035)$ m is selected to match the data available on the TMR website [24]. This point is within the boundary layer of the bump. The present calculations are being compared in this case with results obtained with both the FUN3D and CFL3D codes [24]. The overall tendency of the solution to converge to a common value, as the mesh is refined, is, once again, clearly displayed for all calculations. For the eddy viscosity values at this given point, one can see some differences between the results obtained with the different gradient reconstruction schemes. However, these differences are still fairly small, when compared to the differences between BRU3D results and those from CFL3D and FUN3D. Probably, the largest difference between the BRU3D results is observed when comparing the calculations with $L0E$ and $LJ0$ schemes with mesh 1, but even this difference only accounts for about 2.5 % of the local eddy viscosity coefficient values. Clearly, all differences are reduced as the mesh is refined, as already noted.

Another aspect that can be observed in the results in Figs. 5 and 6, but especially those in the later figure, is that the calculations with FUN3D and CFL3D seem to yield property values, which are closer to those in the mesh converged solution, even at coarser meshes. The authors have performed some preliminary evaluations of this issue and it seems that the behavior is related to the different approaches used for computing the cell-averaged gradient value among the three codes. As indicated in the paper, BRU3D calculates the cell-averaged gradients using a Green-Gauss formulation, whereas FUN3D, for instance, uses a least-squares approach for computing the gradients employed in the linear property reconstruction. Unfortunately, the current version of BRU3D is implemented in such a way that the use of a least-squares approach would decrease its efficiency considerably. As indicated in the literature [11, 16, 19], the use of the Green-Gauss approach can produce significant errors at interfaces between cells of different geometry, which is particularly critical in coarse meshes. As the mesh gets refined and the cell quality is increased, this error should decrease, which explains the behavior observed in Fig. 6.

Finally, the authors want to address efficiency issues for the calculations of the present test case. Although steady state solutions appear to be identical among all three interface gradient reconstruction schemes when mesh 3 is considered, the convergence process for the $L00$ case behaves considerably different than those for the other two cases. Figure 7 shows the convergence histories for the $L_\infty$ norm of the residue for all six equations, obtained with each scheme, for the calculations with mesh 3. It is clear that all calculations reach machine zero. However, the computations with both the $L0E$ and $LJ0$ schemes achieve machine zero convergence in approximately 1,500 iterations. In contrast, the $L00$ solution requires more than 10,000 iterations to be fully converged. Thus, the $L00$ scheme effectively becomes more expensive, in terms of computational costs, than the other two schemes when the system steady state is sought. All calculations used the maximum CFL number allowed by the process described in Sec. 2.3.

# 4 NASA CRM-HL Multielement Airfoil

## 4.1 Geometry, Meshes and Flow Conditions

The NASA Common Research Model High-Lift (CRM-HL) multielement airfoil is a 2-D configuration generated from the complete 3-D NASA CRM-HL configuration [35]. The geometry and the flow conditions that define this test case were developed by the AIAA Geometry and Mesh Generation Workshop (GMGW) committee [34]. The flight conditions simulated in the present work follow the problem setup established for the special session entitled "Mesh Effects for CFD Solutions", held at the virtual AIAA Aviation 2020 Forum.

The meshes for this test case are available at the TMR website [24]. The present study considers three hybrid meshes from the group of computational grids labelled "Family 1" category of grids, according to the nomenclature of the website [24]. The "Family 1" meshes comprise a sequence of 7 unstructured grids, with different levels of refinement. These grids are formed with, mostly, quadrilateral cells, mainly clustered around the airfoil surface, and with a few sparse triangles for geometric compatibility. There is particular interest in observing if interfaces between quadrilateral and triangular cells would yield any additional discretization errors. The specific meshes used in the present calculations, among those available at TMR as the "Family 1" meshes, were grids L2, L5 and L7. These meshes have, respectively, 294,552, 1,680,220, and 5,980,952 cells. An overview of the L5 mesh is presented in Fig. 8, with focus on the area near the airfoil surface. Zoomed-in views of the regions surrounding the slat and the flap cove are also shown in the same figure.

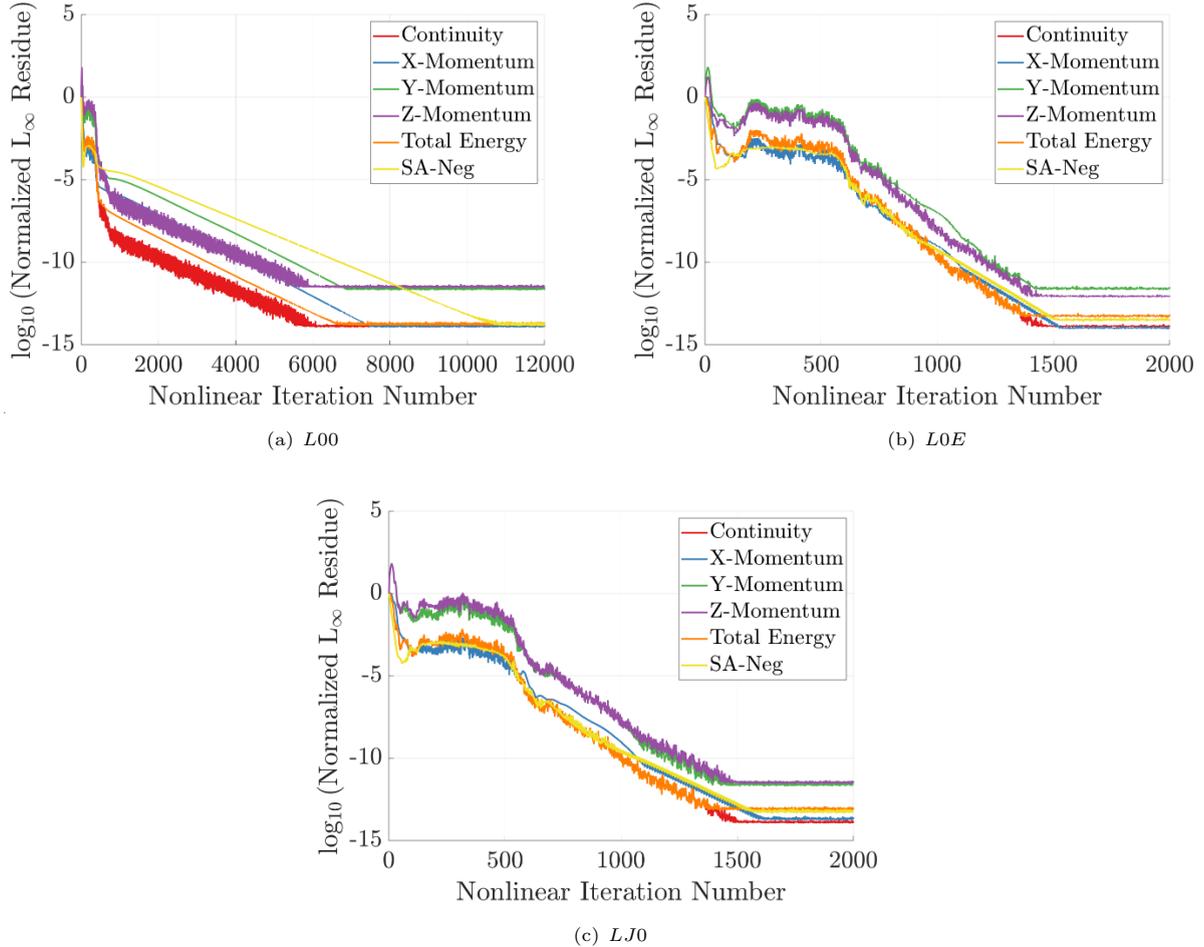

(a) *L00*

(b) *L0E*

(c) *LJ0*

**Fig. 7** Residue convergence history of the bump-in-channel case using mesh 3 and different gradient reconstruction schemes.

It should be observed that the meshes under the "Family 1" category [24] are constructed such that the wall proximity is the primary factor controlling grid clustering. Hence, there is really no attempt to provide additional clustering in regions of the domain where wakes are expected to develop. Moreover, the present code, BRU3D, can only handle a strictly three-dimensional formulation. Therefore, the calculation of the flow over 2-D configurations, as the present one, are handled by extruding one computational cell in the spanwise direction. Hence, the actual computational mesh is composed of hexahedral and triangular-based prism cells.

The usual approach for handling external flows is also used in the present test case. Hence, the simulation has wall boundary conditions and farfield conditions. Non-slip adiabatic wall conditions are assumed at the airfoil surface, and non-reflective farfield boundary conditions, based on Riemann invariants, are used at the farfield boundary. The freestream conditions consider that the freestream Mach number is given by $M_\infty = 0.2$ and the freestream Reynolds number is $Re_\infty = 5 \times 10^6$. The freestream Reynolds number is based on the reference chord $c_{ref} = 1$ m, and on the freestream static temperature $T_\infty = 272.1$ K. The calculations are performed for a 16 deg. angle of attack case. The present study also uses data available at the NASA TMR website [24], from calculations of the flow over the same configuration and flight condition using the FUN3D code [24], for the comparison of the results here obtained.

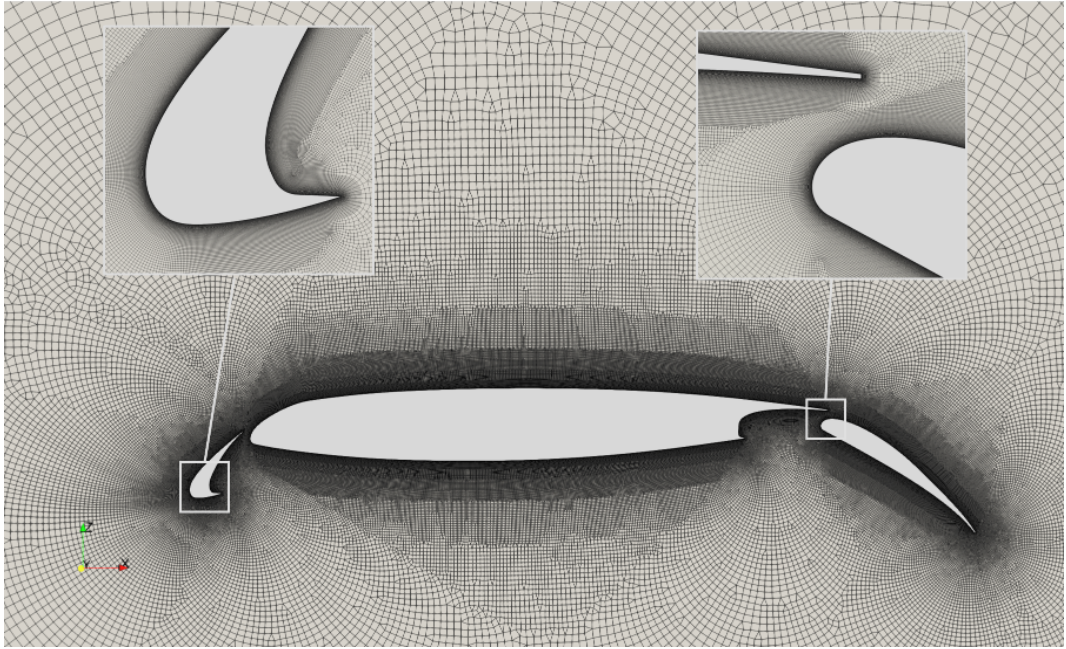

**Fig. 8** Overview of the L5 computational mesh used in the NASA CRM-HL multielement airfoil case. Focus is given to the region surrounding the airfoil surface, and to zoomed-in views of the slat and flap cove areas.

### 4.2 Convergence Rates and Global CFL Behavior

The calculations of the flow over the NASA CRM-HL multielement airfoil are used to demonstrate the type of convergence rates which can be achieved with the present formulation. The results in the present investigation are considered to be properly converged to steady state when a decrease of at least 10 orders of magnitude is observed in the $L_\infty$ norm of the residue of the continuity equation. As an example of the type of behavior that was typically observed in the present study, for the cases that did converge to steady state, Fig. 9 presents the $L_\infty$ norm of the residue for all equations for this CRM-HL airfoil test case. These calculations were performed with the L2 mesh and using the $L0E$ interface gradient reconstruction scheme. Residues shown in the figure are normalized by their first iteration value.

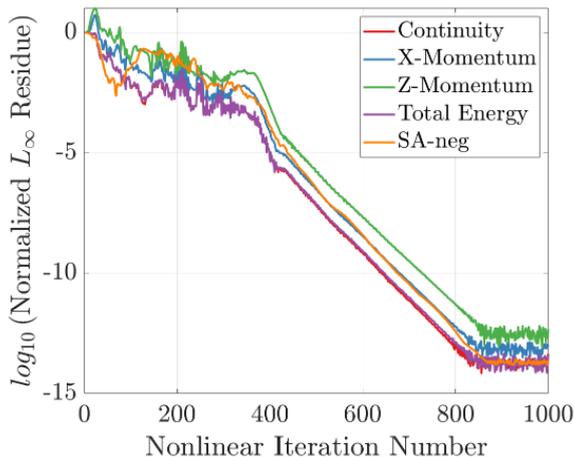

**Fig. 9** Normalized $L_\infty$ residue, as a function of the number of nonlinear iterations, using mesh L2 of the CRM-HL multielement airfoil case.

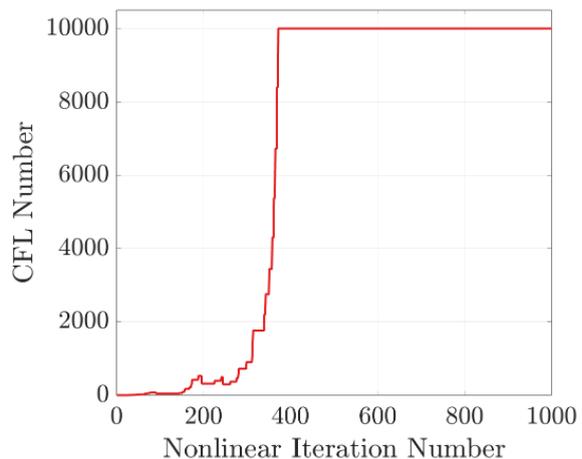

**Fig. 10** Global CFL value assumed throughout the simulation of the CRM-HL multielement airfoil with mesh L2.

Figure 10 shows the behavior of the global CFL number for the same calculation, that is, the simulation for which the residue histories are shown in Fig. 9. As previously indicated, the global CFL number is dynamically controlled throughout the simulation, as described in Sec. 2.3. For all computations reported in the present work, the initial value of the CFL number is 0.1, and we put a cap on the maximum CFL number allowed at 10,000. It is clear that the residue control algorithm successfully keeps the CFL number to lower values during the initial solution transient, without any user input. Then, as an approximately monotone residue convergence is achieved, the algorithm rapidly increases the CFL number, up to the maximum value allowed, yielding a very efficient machine zero convergence process. Obviously, additional tuning of the residue control algorithm parameters could also be done, in order to further improve the convergence rate. The authors, however, are very satisfied with the present performance of this algorithm.

It is interesting to also observe the convergence behavior of the simulation in terms of integrated quantitites. Figure 11 presents values for the lift and drag coefficients, $C_L$ and $C_D$, for the same test case, *i.e.*, flow over the CRM-HL multielement airfoil at $M_\infty = 0.2$, $Re_\infty = 5 \times 10^6$, and $\alpha = 16$ deg., computed with the L2 mesh and using the $L0E$ scheme for interface gradient calculations. The computed aerodynamic coefficients converge up to five significant figures after about 500 iterations, at which point the residues associated to all equations have decreased by, at least, 6 orders of magnitude, as indicated in Fig. 9. Therefore, it is clear that the steady state condition is indeed being achieved. Other cases discussed in the present paper also follow the same overall convergence pattern, unless otherwise stated.

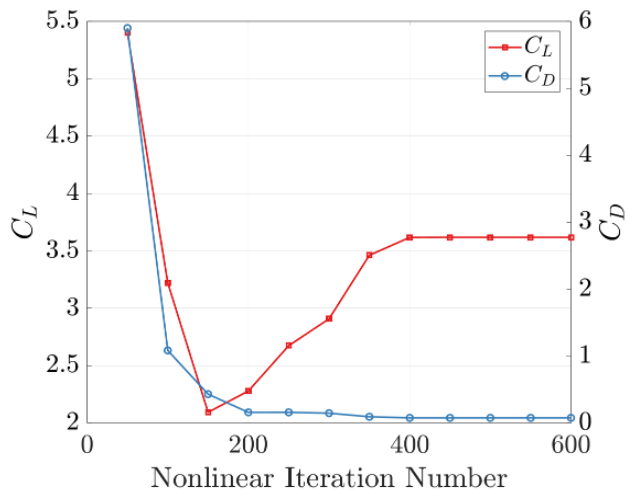

**Fig. 11**  Lift and drag coefficients, as a function of the nonlinear iterations, for the CRM-HL multielement airfoil case computed using the $L0E$ scheme with mesh L2.

### 4.3 Sensitivity to Control Parameters

The formulation implemented in the BRU3D code has, at least, two parameters, that are part of the mathematical formulation in use, but which are rather arbitrary and, at the end of the day, are purely numerical. Therefore, the present effort also performs a brief study of the sensitivity of the computational results to these control parameters. These parameters are present in the formulation of the entropy fix addition to the Roe matrix eigenvalues [32, 38, 39], and in the limiter formulation [20, 31]. They are denoted $\epsilon_H$ and $\epsilon_W$, respectively, in the formulation previously presented. The study considers the same CRM-HL airfoil test case, computed in the L2 mesh and using the $L0E$ interface gradient calculation scheme.

Figure 12 presents an indication of the dependency of the lift and drag aerodynamic coefficients, calculated for this test case, on the value of the entropy fix control parameter, $\epsilon_H$. As previously indicated, the literature [38, 39] suggests that this parameter should be kept within the $0 < \epsilon_H \leq 0.25$ interval. The plots in Fig. 12 indicate that, within this interval, the maximum change in the drag coefficient is of 9 drag counts for a high-lift condition. This is not a negligible variation in drag, but it represents about 1 % of the total drag coefficient at this flight condition. The changes in the lift coefficient are even smaller, amounting to only about 0.4 % of the calculated $C_L$ values. Therefore, the sensitivity of the results to this parameter is not really relevant. Hence, all the calculations in the present paper have used a conservative value of $\epsilon_H$ of 0.05. This has been a good compromise that does not degrade the quality of the results, while it maintains suitable values of artificial dissipation, contributing to the stability of the numerical method.

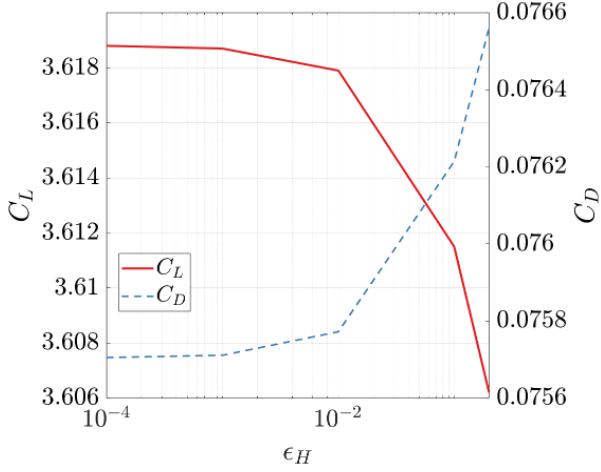
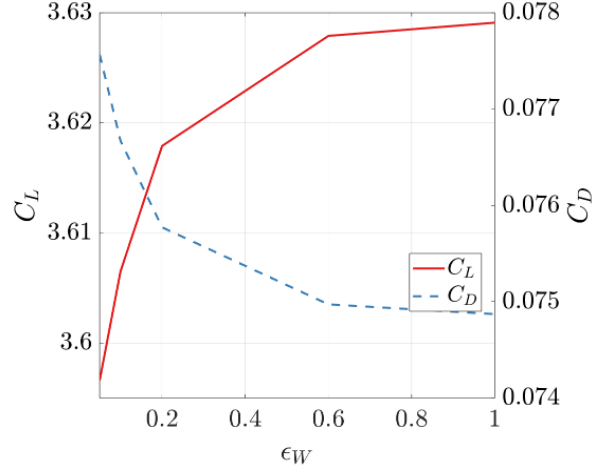

**Fig. 12** Sensitivity of the predicted CRM-HL multielement airfoil aerodynamic coefficients to the $\epsilon_H$ parameter when using the L2 mesh. The horizontal axis is in *log* scale.

**Fig. 13** Sensitivity of the predicted CRM-HL multielement airfoil aerodynamic coefficients to the $\epsilon_W$ parameter when using the L2 mesh.

The work also performed an analysis of the sensitivity of the results to the value of the limiter control parameter, $\epsilon_W$. Similar to the previous discussion, Fig. 13 presents values of lift and drag coefficients for a range of $\epsilon_W$ values. As indicated in the literature [20], the recommended interval for this parameter is $0.01 \leq \epsilon_W \leq 0.20$. Therefore, if one considers only the values of the aerodynamic coefficients obtained for this range of $\epsilon_W$, it is possible to observe in Fig. 13 a variation in the drag coefficient of the order of 18 drag counts, which is about twice as much effect as we have observed in the previous discussion. The variation in $C_L$ is slightly larger than in the previous discussion, but still of the order of 0.55 %, which is quite small.

Although this variation in the predicted lift coefficient is not significant, the change in the computed drag coefficient might start to impact the results. An analysis of the actual flowfield results at the converged solutions indicated that, for subsonic cases, the limiter is more inclined to activate in regions where geometrical discontinuities are present, such as sharp edges. In the present case, this translates to a very small region near the trailing edge of each airfoil component, where the limiter assumes a value different from 1. By increasing the value of $\epsilon_W$, the number of affected cells decreases, which reduces the amount of artificial dissipation that is introduced to the aforementioned regions. The reduction in the number of affected cells, in turn, improves the solution of all trailing edge wakes, resulting in different values for the predicted aerodynamic coefficients. This effect should become less pronounced with each successive mesh refinement due to the nature of the intrinsically added dissipation of an upwind scheme. With all those considerations in mind, for the current study, a value of $\epsilon_W = 0.08$ is adopted, since it seems to work well for all cases studied here.

### 4.4 Effects of the Gradient Reconstruction Schemes

In order to give a general overview of the flowfield which is being calculated for this test case, Fig. 14 presents a visualization of Mach number contours superimposed by flow streamlines. The calculation with results shown in Fig. 14 was performed using the L2 mesh and the $L0E$ interface gradient reconstruction scheme. As already indicated, this is a high-lift flow condition. There are well-defined recirculation zones on the slat lower surface and on the flap cove regions. The wake regions are also clearly defined over the airfoil main element and in the region downstream of the flap. In other words, one is seeing the overall flow features expected for such a high-lift flow condition.

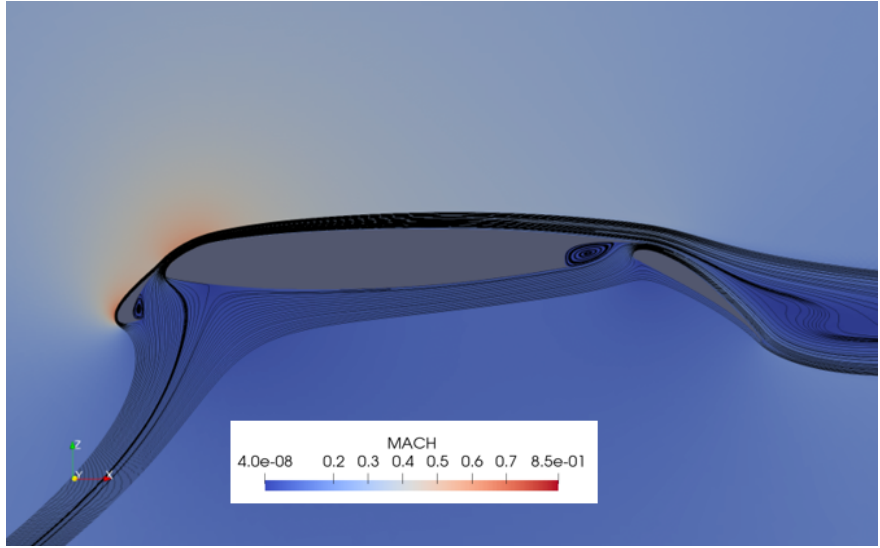

**Fig. 14** Streamlines for the CRM-HL case, superimposed to Mach number contours, obtained using the *L0E* scheme and the L2 mesh.

A comparison of the effects of the different interface gradient calculation schemes on the lift and drag coefficients is shown in Figs. 15 and 16 as a function of mesh refinement. As already indicated, for all computations in the present work, the cell-averaged gradients are calculated using the Green-Gauss formulation. One can observe that the calculations with the *L00* and *L0E* interface gradient schemes yield essentially coincident results for both aerodynamic coefficients in this test case. Results with the *LJ0* scheme display slight differences, especially for the coarser meshes. As the mesh is refined, all computations tend to converge to the same values of lift and drag coefficients. It must be emphasized, however, that we have no results for the calculation with the *L00* scheme on the L7 mesh, *i.e.*, the finest mesh, since the numerical method becomes unstable for this fine mesh under the present flow conditions. This is a pattern observed in the present effort, that is, when we are able to converge the calculations with the *L00* scheme, the results are very similar to those obtained with the *L0E* scheme. However, quite a few calculations do not remain stable using the *L00* scheme. The next section will explore this issue further, but the absence of sufficient artificial dissipation to suppress the decoupling problem that is inherent to the *L00* formulation becomes evident in the present results.

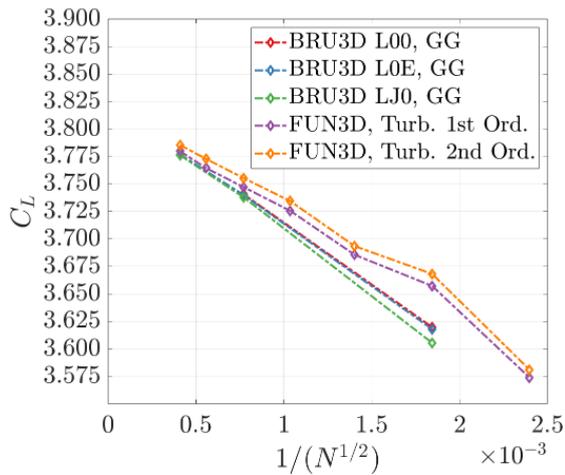

**Fig. 15** Lift coefficient values, as a function of the mesh size, for the CRM-HL airfoil case.

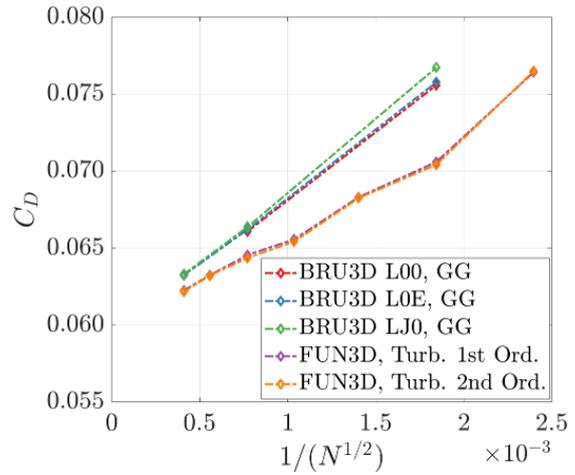

**Fig. 16** Drag coefficient values, as a function of the mesh size, for the CRM-HL airfoil case.

The present results are compared with data available in the literature [24]. In particular, for the lift and drag calculations, our results are compared to FUND3D calculations using the same family of meshes. These results are available in the TMR website [24], which actually reports FUND3D results computed with the SA-neg turbulence model with two different spatial discretizations for the advection terms, that is, both 1st- and 2nd-order accurate results are reported at TMR [24]. It is important to emphasize that all terms in the SA-neg equation are discretized with 2nd-order spatial accuracy in our calculations. One can observe in Fig. 15 that there are slight differences in the lift coefficients between the results obtained with the two versions of the FUN3D code. These differences are, however, smaller than those observed between the present results and those reported at TMR. The differences in the lift coefficient, however, are fairly small, amounting to less than 2 % even at the coarsest mesh (L2) at which we performed calculations for this test case. Nevertheless, as the mesh is refined, all lift coefficient results tend to approach each other, and the differences are very minor, less than 0.2 %, in the L7 mesh results.

The behavior observed for the drag coefficient results, shown in Fig. 16, is somewhat similar to what we just described for the lift coefficient, except that the differences in drag coefficient values are considerably larger, especially at the coarse mesh. For the L2 mesh calculation, the largest difference between our results and those from FUN3D is of the order of 65 drag counts, *i.e.*, about 8 %, which is significant even for a high-lift calculation. As indicated, these differences are considerably reduced with mesh refinement. For the L7 mesh calculation, our calculations yield results with about only 1.5 % difference with regard to the FUN3D computations [24]. It is clear, however, that none of the calculations presented here, for this test case, have reached the asymptotic convergence region for both lift and drag coefficients.

Figure 17 presents the pressure coefficient, $C_p$, distributions along the airfoil surface for the three interface gradient reconstruction schemes under consideration here. Detailed views of the main element and flap suction peaks are also indicated in this figure. Our results shown in Fig. 17(a) are calculated using the L5 mesh, since this is the finest mesh for which we obtain converged solutions with all three gradient schemes. The present calculations are compared to results obtained using the FUN3D code [24], using the L7 mesh and 2nd-order spatial discretization for the turbulence model equation. All results are very similar throughout most of the airfoil surface. One can observe a few differences at the main element and flap suction peaks. However, even at those points, as we can in the insets in Fig. 17, differences are small. One can see that the present computations with the $L00$ and $L0E$ interface gradient calculation schemes are essentially indistinguishable, whereas the computations with the $LJ0$ scheme yield slightly different results. The differences between the present $C_p$ results and those calculated with FUN3D are slightly larger but remain small, of the order of 1 % and 3 %, respectively, at the main element and flap suction peaks. In any event, all of these differences are indeed very minor and, since they are very localized, they are probably not significant from an engineering point of view, especially considering that different meshes were used for the present and the reference results. Furthermore, the differences among the results with the three gradient schemes here discussed are even smaller and, hence, the stability properties of the method probably are the sole reason for considering one scheme over the other for the present application. Figure 17(b) shows the converged solution using the L7 mesh. As expected, the difference between the solutions decrease even further when the mesh is refined.

A visualization of velocity component profiles, plotted along the vertical, $z$, direction for three chordwise stations, is presented in Fig. 18. The stations are selected such that the 1st one, at $x = -0.03$, is located over the slat region; the 2nd station, at $x = 0.4$, is in the main airfoil region; and the last station is over the flap, at $x = 0.95$. The figures show the longitudinal and vertical velocity components, $u$ and $w$, respectively, normalized by the magnitude of the freestream velocity, $U_{ref}$. Similar to the results reported in the previous paragraph, the current results are calculated in the L5 mesh, since this is the mesh for which we have results with all three interface gradient calculation schemes. The present results are again compared to FUN3D calculations, available at the TMR website [24], using the L7 mesh. As one can see in Fig. 18, all velocity component profiles over the slat and the main airfoil element are essentially identical. For the flap profile, one can see that the present calculations also yield essentially identical profiles, but there are slight differences between the profiles obtained in our calculations and the data available at TMRe [24]. The trends are absolutely identical, but there are very small differences in the values of the velocity components for the flap profile.

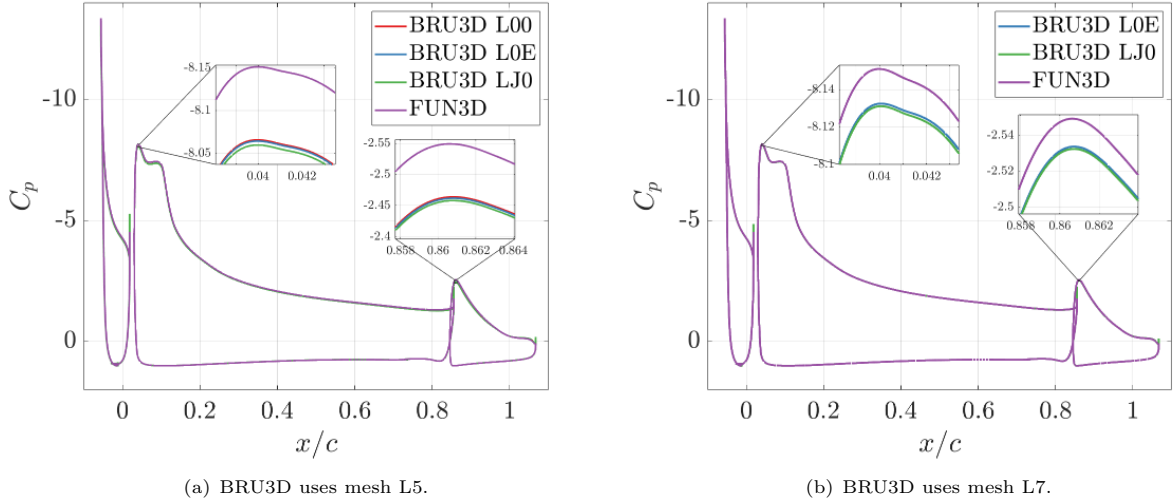

(a) BRU3D uses mesh L5.   (b) BRU3D uses mesh L7.

**Fig. 17** Pressure coefficient distributions for the CRM-HL multielement airfoil case. FUN3D calculations use the L7 mesh.

## 5 Transonic Flow over the ONERA M6 Wing

### 5.1 Configuration Definition and Flight Conditions

The ONERA M6 wing [36] is a classical test case used for verification and validation of many CFD codes under transonic conditions. The authors have used the geometry specifications and test case flight conditions as defined in the TMR website [24]. A family of three meshes has been developed, namely, the coarse, medium and fine meshes. These grids are generated using a mesh generation software available at TMR [24], and they are configured with a blunt trailing edge. These coarse, medium and fine grids have, respectively, 1.18, 9.44, and 15.73 million cells, approximately. An overview of the medium mesh can be seen in Fig. 19, which presents an overall view of the region around the upper surface of the wing and more detailed views of the two regions indicated. The insets focus on the leading and trailing edge regions of the wing tip.

The meshes are composed of hexahedral and tetrahedral cells, as one might infer from the surface grids shown in Fig. 19. The tetrahedral cells are used solely on the wing tip region, growing from the wing surface until the farfield is reached. Moreover, there are no prism layers being used in this case. For the remainder of the computational domain, hexahedral cells are used and, as expected, the hexahedra grow from the wing surface to the farfield boundary surface. Therefore, there is a clear interface between tetrahedral and hexahedral cells near the wing tip region, which is also a region where high property gradients are expected to develop due to the presence of a wing tip vortex.

This is a transonic test case with freestream Mach number $M_\infty = 0.84$. The reference Reynolds number, based on the wing root chord, is $Re_{c_{root}} = 14.6 \times 10^6$. The wing root chord is $c_{root} = 1$ m. The reference freestream temperature, $T_\infty$, is 300 K, and the flight angle of attack is 3.06 deg. The test case setup uses the typical boundary conditions for aerodynamic external flows, i.e., no-slip adiabatic boundary conditions on the wing surface and non-reflective farfield conditions, based on Riemann invariants, over the domain outer surface. Moreover, a symmetry plane is assumed at the wing root, hence, reducing the computational costs by only simulation one-half of the wing.

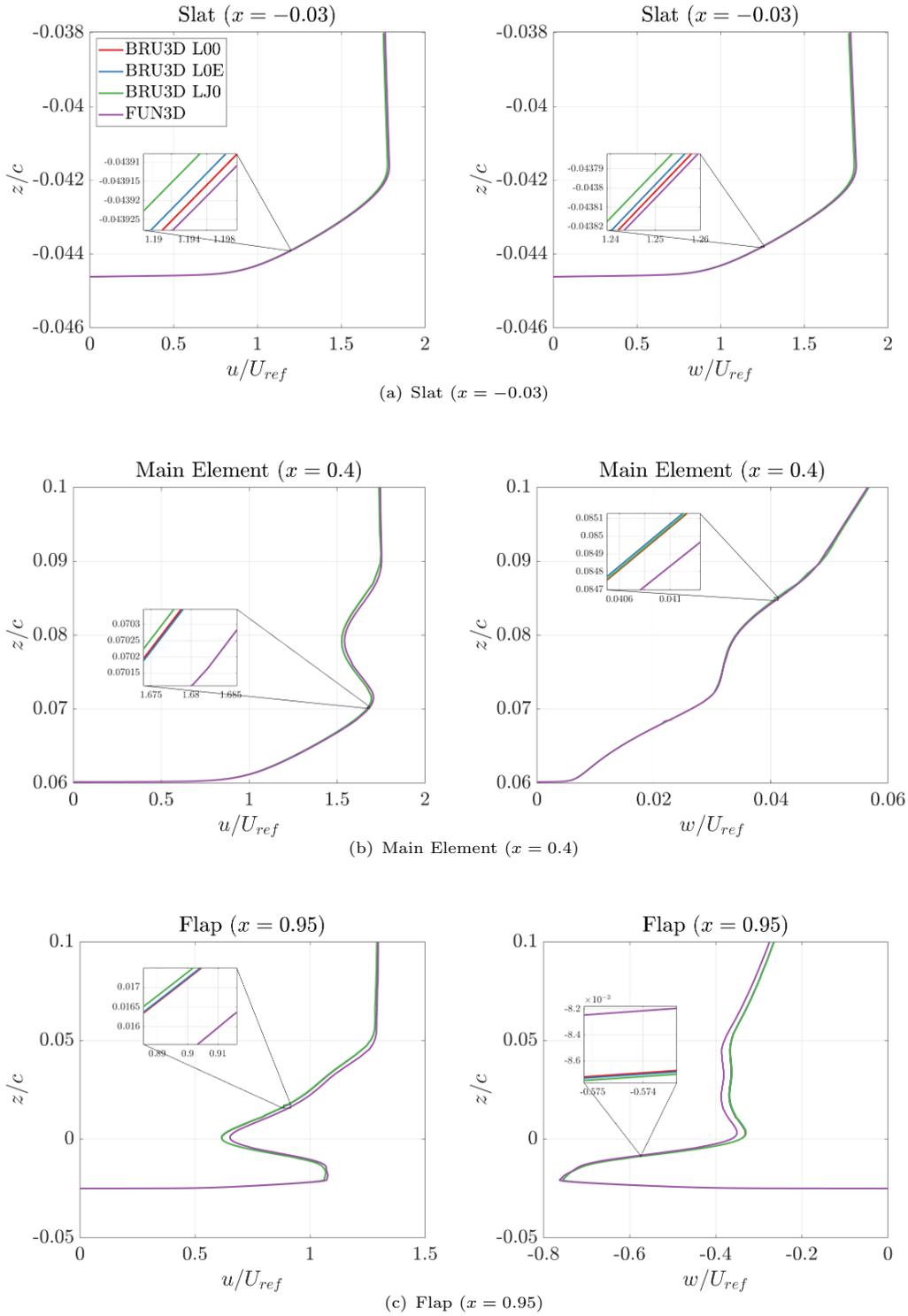

**Fig. 18** Velocity profiles, evaluated at three different chordwise locations, for the CRM-HL case with mesh L5. Data from FUN3D have used mesh L7.

## 5.2 Results and Discussion

The ONERA M6 wing at this flight condition develops a very interesting shock structure over the wing upper surface. In the inboard portions of the wing, there are two shocks along the chord, an upstream shock and a downstream shock. As one moves towards the wing tip, these two shock structures merge into a single shock wave over the wing surface. Hence, this gives the impression of a $\lambda$ shock structure. Figure 20 presents the pressure coefficient contours over the wing upper surface calculated with the present code for this flight condition. The calculation that led to the results shown in Fig. 20 used the fine mesh and the $L0E$ interface gradient calculation scheme. The $\lambda$ shock structure, previously described, is very well defined and clearly visible in the figure.

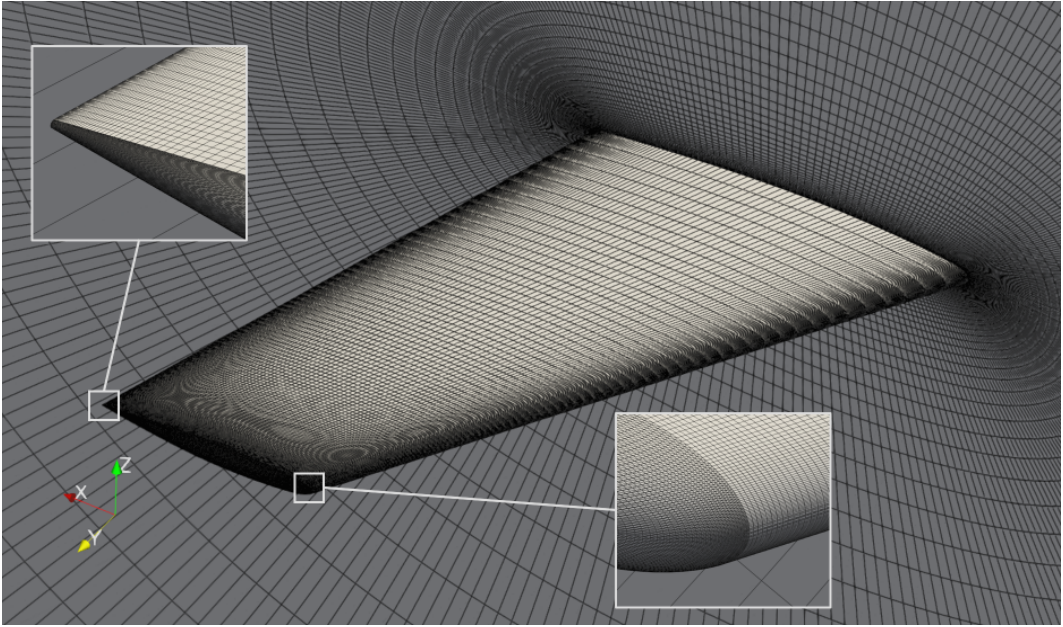

**Fig. 19** Overview of the medium mesh used in the ONERA M6 case. Focus is given to the region surrounding the wing surface. Zoomed-in views of the leading edge and trailing edge regions of the wing tip are also shown.

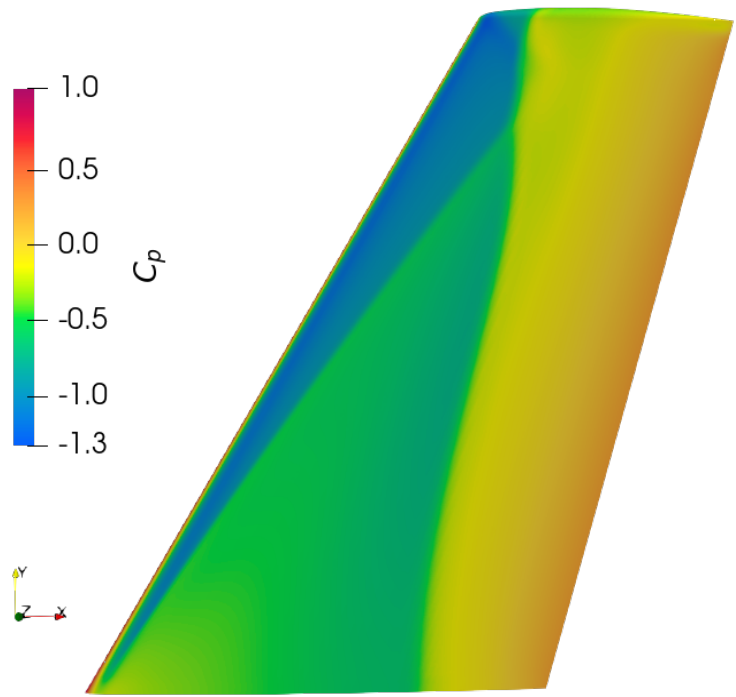

**Fig. 20** Pressure coefficient contours over the upper surface of the transonic ONERA M6 wing, computed using BRU3D with the $L0E$ scheme and the fine mesh.

The authors have not been able to obtain stable solutions with the $L00$ interface gradient calculation scheme, for this transonic flight condition, for any of the three meshes used here. The high-frequency errors that may develop in the solution, due to the uncoupling present in the $L00$ interface gradient calculation procedure [7], are severely present in this test case. Hence, all calculations with the $L00$ scheme have quickly diverged for the three meshes used. Therefore, only results with the $L0E$ and $LJ0$ gradient schemes are presented and discussed. Figure 21 presents pressure coefficient distributions at 6 spanwise stations along the semi-span, namely, for $\eta = 0.20$, 0.44, 0.65, 0.80, 0.90, and 0.99. The results are shown for all three meshes, but only for the calculations with the $L0E$ gradient scheme, because the results with the $LJ0$ scheme were essentially identical in each corresponding mesh. For comparison purposes, results obtained with the CFL3D code and experimental data [24, 36] are also shown in the figure. The CFL3D results are also available in the TMR website [24], and they were calculated with a much finer mesh with approximately 69.2 million cells.

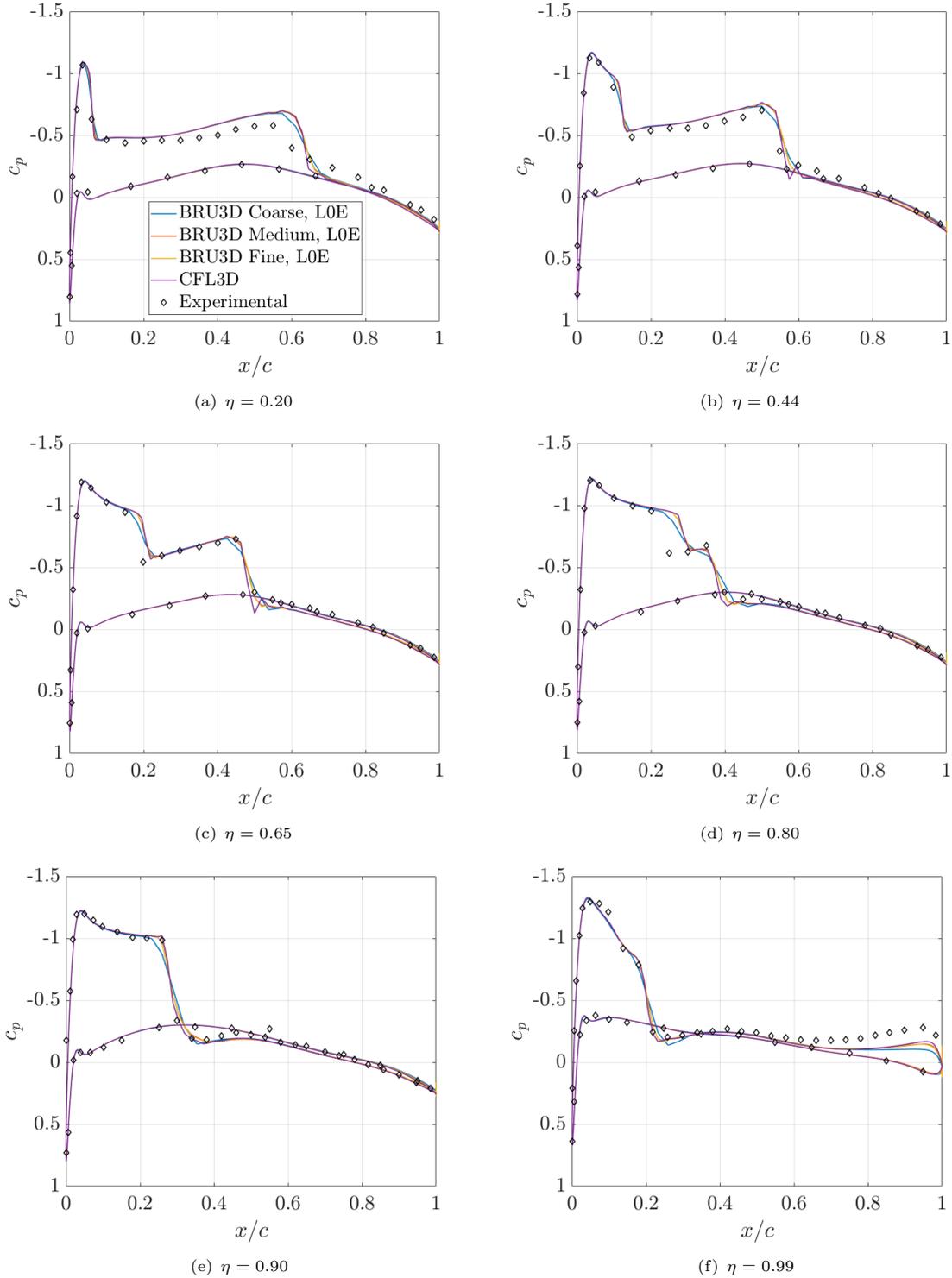

**Fig. 21** Pressure coefficient distributions over the surface of the transonic ONERA M6 wing for span stations, $\eta$. Data from CFL3D [24] and experiments [24, 36] are also presented.

One can observe in Fig. 21 that there are no significant differences in the $C_p$ distributions between the present calculations with the medium and fine meshes. On the other hand, one can clearly see the differences for the present calculations with the coarse grid, when compared to the results with the other two grids, especially at the shock locations. As expected, results with the coarse mesh are fairly diffused near the shock waves. There are also differences in the coarse mesh results at the 99 % station, in the suction region due to the effect of the wing tip vortex. The present calculations with both medium and fine meshes also present a very good agreement with the CLF3D results, even despite the fact that those results were computed in a much finer grid. In general, there is an overall good agreement between the numerical simulations and the experimental data. Exceptions for this statement could, however, be observed at the shock locations, at the $C_p$ plateau between the two shocks for the two innermost stations, and again at the suction region near the trailing edge at the $\eta = 0.99$ station.

With regard to the shock location, the largest differences between computation and experiment seem to occur at the $\eta = 0.80$ station. Although there is a very good match of the results for the downstream shock, the position of the upstream shock predicted by all computations, including the reference CFL3D calculations [24], is considerably further downstream than indicated by the experimental data. This seems to indicate that the two shocks, in the computations, are merging slightly inboard when compared to the experiments. Finally, it is also very clear from the results at the $\eta = 0.99$ station that the calculations cannot adequately represent the additional suction created by the wing tip vortex. It is true that mesh refinement, for the present calculations, has slightly improved the results in this region, but there are still non-negligible differences.

Table 1 presents lift and drag coefficients, calculated with the present code with both medium and fine grids. The results are shown for both $L0E$ and $LJ0$ interface gradient calculation schemes, and they are compared to aerodynamic coefficients obtained with CFL3D with the much finer grid previously reported. One can observe in the table that our results obtained with the $L0E$ and $LJ0$ interface gradient calculation schemes, for each mesh level, are very close to each other both for $C_L$ and for $C_D$. For instance, even in the medium mesh, the difference in $C_D$ is less than 1 drag count. The percent differences indicated in the last two columns in Tab. 1 refer to the comparison with the corresponding CFL3D data. As one can see, the largest difference occurs in the drag coefficient calculated with the present code in the medium mesh and with the $LJ0$ gradient scheme. However, this difference is only about 3 %, and our medium mesh has more than 7 times fewer cells than the grid used for the CFL3D calculation. The differences in $C_D$ results consistently decrease with mesh refinement, although the variations are not terribly significant with the meshes used in the present calculations.

**Table 1** Aerodynamic coefficients predicted by each scheme. Results from CFL3D are obtained from the TMR website [24].

| Source | $N$ (cells) | $C_L$ | $C_D$ | $C_L$ Diff. | $C_D$ Diff. |
| --- | --- | --- | --- | --- | --- |
| BRU3D $L00$ | 9,437,184 | Diverges | Diverges | - | - |
| BRU3D $L0E$ | 9,437,184 | 0.27414 | 0.017414 | +1.79% | +2.70% |
| BRU3D $LJ0$ | 9,437,184 | 0.27375 | 0.017478 | +1.64% | +3.08% |
| BRU3D $L0E$ | 15,728,640 | 0.27469 | 0.017409 | +1.99% | +2.67% |
| BRU3D $LJ0$ | 15,728,640 | 0.27438 | 0.017384 | +1.88% | +2.52% |
| CFL3D | 69,206,016 | 0.26932 | 0.016956 | Reference | Reference |

Figure 22 presents skin friction coefficient distributions, computed in the $x$-direction, for the $\eta = 0.96$ station along the semi-span. The results shown in the figure are calculated with both the $L0E$ and $LJ0$ interface gradient calculation schemes in the medium mesh.

The reference CFL3D results were calculated in the very fine grid indicated in Tab. 1. Although one can see that there are differences among the various calculations, such differences are indeed very minor. Moreover, the overall trends of all $C_f$ curves are essentially identical. With the objective of highlighting some of the differences in $C_f$, Fig. 22 has two insets, which offer additional details of the $C_f$ distributions over the airfoil. In general, throughout most of the airfoil surface, our results with the $L0E$ and $LJ0$ gradient schemes are closer to each other, and there are larger differences with regard to the reference CFL3D data. However, there are regions in the $C_f$ curves in which the differences between our calculations with the $L0E$ gradient scheme and the reference CFL3D data are smaller than the differences between our own calculations.

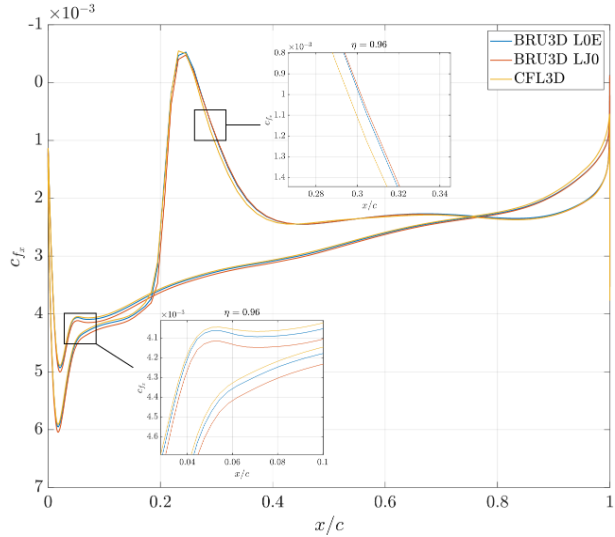

**Fig. 22** Skin friction coefficients in the $x$-direction at the $\eta = 0.96$ station for the transonic ONERA M6 wing case, computed using BRU3D with the medium mesh and CFL3D with the very fine mesh.

# 6 Concluding Remarks

The present work has addressed some aspects which are relevant for the accurate and stable computation of compressible, high Reynolds number flows. The study was performed in the context of a general unstructured grid, 2nd-order, finite volume method. The main aspect studied was the effect of different cell-interface gradient reconstruction techniques, which are necessary in order to compute the viscous terms in the context of such a numerical method. Three different forms of computing these interface gradients were implemented, and these were termed $L00$, $L0E$, and $LJ0$ schemes. For many of the test cases addressed, particularly for all ONERA M6 wing cases and for the fine grid calculation of the NASA CRM-HL multielement airfoil, the simplest scheme, $L00$, in which the interface gradient is essentially a weighted-average of the cell-averaged gradients, does not yield stable simulations. For the third test case here discussed, namely the subsonic bump-in-channel flow, which was solved using highly orthogonal meshes, the $L00$ scheme produced results as good as the ones obtained with the other schemes, but it increases the required number of iterations to achieve steady state by almost 7 times when compared to the other interface gradient calculation methods.

On the other hand, no significant differences were observed on the results calculated with either the $L0E$ or the $LJ0$ schemes for all test cases studied. The $LJ0$ scheme is known in the literature to be capable of achieving up to fourth-order accuracy in terms of spatial discretization error when the appropriate conditions are met. However, for cases with complex geometries subject to advection dominated flows, such as the ones studied here, the $LJ0$ scheme is held back by the discretization scheme used in the treatment of the inviscid fluxes, which in this study is limited to second-order accuracy. Thus, no practical differences are observed between the $L0E$ or the $LJ0$ schemes in the test cases considered in this research, which were chosen due to their capability of representing real aerospace CFD applications. In order to be complete in the analysis, it must be said that the results obtained with these two interface gradient calculation schemes are not always identical, but the differences observed are not really relevant. Therefore, based on the present results, the authors cannot suggest the use of one scheme over the other. Both in terms of quality of the solution and in terms of computational costs, the $L0E$ and $LJ0$ gradient reconstruction schemes have yielded essentially equivalent results.

The present study has also performed a sensitivity analysis with regard to some numerical control parameters which exist in the present formulation. Our results have indicated that the present formulation seems to be highly insensitive to the entropy fix control parameter, $\epsilon_H$, provided that its value is kept within the range suggested in the literature. On the other hand, the results do indicate some sensitivity to the parameter, $\epsilon_W$, present in the limiter formulation, even for a subsonic high-lift flow. Although detailed results are not shown in the paper, the authors have experimented with three different limiter formulations. All formulations have some numerical control parameter, and they all exhibit some effect of this parameter in the results. The study observed that, for subsonic test cases, the limiter is more likely to activate in regions where geometrical discontinuities are present, such as sharp edges. Hence, an increase in the control parameter tends to reduce the number of cells actually affected by the limiter, which, in turn, reduces the amount of artificial dissipation that is introduced in those flow regions. The limiter effect on the aerodynamic solutions seems to decrease with mesh refinement.

Finally, the present paper has proposed a novel strategy for driving convergence to steady state solutions by dynamically controlling the simulation CFL number based on the current residue behavior. The proposed procedure, coupled to the GMRES linear system solver and to adequate definition of flux Jacobian matrices, is very robust, computationally inexpensive, and effective in driving residues to machine zero. Obviously, the proposed steady state driver cannot always compensate for instabilities arising due to the uncoupling in the interface gradient calculation scheme, as observed in many situations with the $L00$ scheme in the results here discussed. However, for the calculations that remain stable, it has shown to consistently yield machine zero converged results. The present analysis, however, is completely constrained to the use of a single turbulence model. Thus, a natural extension of the present work is to verify whether the behavior observed here still holds if more feature-rich, and numerically stiff, turbulence models are used.

## Acknowledgments

The authors wish to express their gratitude to the São Paulo Research Foundation, FAPESP, which has supported the present research under the Research Grants No. 2013/07375-0 and No. 2021/00147-8. The authors also gratefully acknowledge the support for the present research provided by Conselho Nacional de Desenvolvimento Científico e Tecnológico, CNPq, under the Research Grant No. 315411/2023-6. The work is further supported by the computational resources of the Center for Mathematical Sciences Applied to Industry, CeMEAI, also funded by FAPESP under the Research Grant No. 2013/07375-0.